\documentclass[prb,twocolumn,showpacs,floatfix,superscriptaddress]{revtex4-2}

\usepackage[utf8]{inputenc}
\usepackage[T1]{fontenc}
\usepackage[margin=2cm]{geometry}
\usepackage{float}
\usepackage{color}
\usepackage{graphicx}
\usepackage{amsmath}
\usepackage{amssymb}
\usepackage{bbm}
\usepackage{indentfirst}
\usepackage{dcolumn}
\usepackage{soul}
\usepackage{mathtools}

\usepackage{subfigure}
\usepackage{multirow}
\usepackage{setspace}

\usepackage{hyperref}

\begin{document}

\title{Theoretical determination of Gilbert damping in reduced dimensions}

\author{Bal\'{a}zs Nagyfalusi}
\email{nagyfalusi.balazs@ttk.bme.hu}
\affiliation{Institute for Solid State Physics and Optics, HUN-REN Wigner Research Center for Physics, Konkoly-Thege M. \'{u}t 29-33, H-1121 Budapest, Hungary}
\affiliation{Department of Theoretical Physics, Institute of Physics, Budapest University of Technology and Economics, Budafoki \'{u}t 8, H-1111 Budapest, Hungary}
\author{L\'{a}szl\'{o} Szunyogh}
\email{szunyogh.laszlo@ttk.bme.hu}
\affiliation{Department of Theoretical Physics, Institute of Physics, Budapest University of Technology and Economics, Budafoki \'{u}t 8, H-1111 Budapest, Hungary}
\affiliation{HUN-REN-BME Condensed Matter Research Group, Budapest University of Technology and Economics, Budafoki \'{u}t 8, H-1111 Budapest, Hungary}
\author{Kriszti\'an Palot\'as}
\email{palotas.krisztian@wigner.hun-ren.hu}
\affiliation{Institute for Solid State Physics and Optics, HUN-REN Wigner Research Center for Physics, Konkoly-Thege M. \'{u}t 29-33, H-1121 Budapest, Hungary}
\affiliation{Department of Theoretical Physics, Institute of Physics, Budapest University of Technology and Economics, Budafoki \'{u}t 8, H-1111 Budapest, Hungary}

\date{\today}

\begin{abstract}

An \textit{ab initio} scheme based on the Kubo-Greenwood linear response theory of exchange torque correlation is presented to calculate intrinsic Gilbert damping parameters in magnets of reduced dimensions. The method implemented into the real-space Korringa-Kohn-Rostoker (RS-KKR) Greens' function framework enables to obtain diagonal elements of the atomic-site-dependent on-site and non-local Gilbert damping tensor.
Going from the 3D bulk and surfaces of iron and cobalt ferromagnets addressed in our previous work [Phys.~Rev.~B 109, 094417 (2024)], in the present paper monolayers of Fe and Co on (001)- and (111)-oriented Cu, Ag, and Au substrates are studied, and particularly the substrate-dependent trends are compared.
Furthermore, the Gilbert damping parameters are calculated for Fe and Co adatoms and dimers on (001)-oriented substrates. It is investigated how the damping parameter of single adatoms depends on their vertical position. This dependence is quantified in relation to the adatoms' density of states at the Fermi energy showing a non-monotonic behavior. By rotating the spin moment of the adatoms and collinear magnetic dimers, an anisotropic behavior of the damping is revealed. Finally, a significant, three- to ten-times increase of the on-site Gilbert damping is found in antiferromagnetic dimers in comparison to the ferromagnetic ones, whilst the inter-site damping is even more enhanced.
\end{abstract}

\maketitle

\section{Introduction}

The detailed understanding of the energy dissipation in magnetic materials is crucial for designing magnetic device components for future applications. Magnetic dissipation processes are often described by the Landau-Lifshitz-Gilbert (LLG) equation \cite{Landau1935,Gilbert2004}, where the rate of the energy dissipation is measured by the usually scalar Gilbert damping parameter $\alpha$.
The magnitude of $\alpha$ governs the energy dissipation and the speed of relaxation from a metastable state to  equilibrium, thus the Gilbert damping largely influences the efficiency of spintronic devices.
Previous studies showed that the magnetic damping is not a simple scalar but rather a tensorial \cite{fahnle2005,fahnle2006,Ebert2011,Bhattacharjee2012,Dhali_2024} and non-local quantity \cite{Thonig2018,Gilmore2009,Lu2023,lu2024}.
The non-locality of the damping parameter is reflected in the wave-number-dependent energy dissipation in thin films \cite{Li2016}, and the connection between the gradients of the magnetization \cite{Yuan2014} and the damping also validated this idea. Moreover, due to the effect of non-local damping, an increased magnon lifetime was reported \cite{Lu2023}.

The LLG equation describing the atomistic spin-dynamic process with non-local energy dissipation takes the following form \cite{Thonig2018},
\begin{align}
\label{eq:Gilbert_nonloc}
 \frac{\partial \vec{m}_j}{\partial t} = 
   \vec{m}_j \times \left(-\gamma \vec{B}_j^\mathrm{eff} + \sum\limits_k \underline{\underline{\alpha}}_{jk} \frac{1}{m_k} \frac{\partial \vec{m}_k}{\partial t}\right)\,.
\end{align}
where $\vec{m}_j$ is the magnetic moment at site $j$ with the magnitude $m_j=|\vec{m}_j|$, $\gamma$ is the gyromagnetic ratio, and the damping term consists of pairwise contributions of strength $\underline{\underline{\alpha}}_{jk}$.

An important property of the Gilbert damping is its anisotropy with respect to the direction of the magnetization. This phenomenon already appears in bulk systems \cite{fahnle2005,Thonig2018,Miranda2021,bulkcikk}, and it becomes more significant in reduced dimensions \cite{xia2021,Brinker2022}, such as in monolayers or atomic clusters. This anisotropy originates from the spin-orbit coupling (SOC) \cite{xia2021,li2019}, thus it evidently disappears in the absence of the SOC.
A giant anisotropy of the Gilbert damping was found in Co$_{50}$Fe$_{50}$/MgO(100) thin films \cite{li2019,xia2021} with a maximum-minimum ratio of $400\%$, and the same effect was reproduced in other CoFe alloys \cite{wang2023,xu2023}.
Moreover, the effect of the local chemical environment on the Gilbert damping was investigated in real-space clusters embedded into FeCo alloys in Ref.~\cite{lu2024}, and strong correlations between the effective local damping and the amount of Co atoms and the average Co-Co distance in the atomic clusters were established. Note that the LLG equation introduced in Eq.~\eqref{eq:Gilbert_nonloc} can further be extended to include the effect of various spin torques \cite{PhysRevMaterials.8.114404} and magnetic inertia \cite{PhysRevB.110.174430}, which are relevant, for example, for terahertz spin dynamics.

There are many different approaches to calculate the Gilbert damping \cite{Kambersky1970,Kambersky1976,Gilmore2007,brataas2008,Starikov2010,Ebert2011,Mankovsky2013,Thonig2014,Thonig2018,Guimaraes2019}.
In our previous work \cite{bulkcikk} we introduced a real-space computational tool based on the Korringa-Kohn-Rostoker (KKR) formalism of Ref.~\cite{Ebert2011} for the diagonal Cartesian matrix elements of the non-local tensorial Gilbert damping covering the 3D to 0D range of magnetic materials. In order to test our method, some relevant damping properties of ferromagnetic (FM) bulk Fe and Co, and bcc Fe(001) and fcc Co(001) surfaces were studied \cite{bulkcikk}.
Our choice of materials fit well to the tendency that most scientific papers on magnetic damping concentrate on bulk transition metals Fe, Co and Ni \cite{Barati2014,Thonig2018, Gilmore2007,Guimaraes2019} and on their alloys \cite{Mankovsky2013,Ebert2011,lu2024}, with a special interest in the Fe$_{50}$Co$_{50}$ alloy \cite{Miranda2021}, which shows a giant anisotropy in thin film forms \cite{li2019,xia2021}.
Other works on differently alloyed films \cite{wang2023,xu2023} also confirmed this giant Gilbert damping anisotropy phenomenon.
The Gilbert damping of chemically homogeneous magnetic thin films was also studied \cite{Costa2015,Chen2023,Guimaraes2019}, and the questions on how the strength of the damping can be controlled by film thickness \cite{Barati2014} or by alloying \cite{Arora2021,Qaid2024} were also investigated.

Despite these efforts, going towards even lower dimensions, the Gilbert damping of magnetic atomic clusters remained a less explored area. Ref.~\cite{Brinker2022} reports on the decomposition of damping contributions of selected individual magnetic atoms and atomic dimers on a Au(111) surface, providing a detailed microscopic understanding of the origin of the damping tensor, also in terms of SOC. This work also proposes the possibilities of including multi-body contributions to the Gilbert damping, which are ultimately important in case of noncollinear magnetic configurations. The reduced number of magnetic atoms makes possible to investigate the dependence of the damping on the magnetic configuration itself, and the authors of Ref.~\cite{Brinker2022} show that the damping can greatly differ whether the spin moments of the two atoms in a dimer are coupled ferromagnetically or antiferromagnetically.

Taking full advantage of our proposed real-space method for calculating the Gilbert damping tensor components of magnets in reduced dimensions in Ref.~\cite{bulkcikk}, in this paper we are studying magnetic 2D monolayers and 0D adatoms and atomic dimers composed of Fe and Co transition metals.
As substrate host materials we select non-magnetic Cu, Ag and Au in (001) and (111) surface orientations to be able to identify substrate-SOC-related trends in the calculated Gilbert damping values.
Our study on the 2D magnetic monolayers concentrates on the comparison of the obtained Gilbert damping values with those of the FM bulk and surface systems. In particular, we investigate the effects of the broadening ($\eta$), the direction of the magnetization, the surface orientation of the substrate and the effect of the SOC on the magnetic damping.
0D magnetic structures are ideal to study on-site and inter-site dampings in a well-controlled manner, similarly to Ref.~\cite{Brinker2022}, where adatoms and dimers were considered. For adatoms, we concentrate on the relation between the damping and their density of states (DOS) through varying their vertical position. The dependence of the damping on the alignment of the atomic spins in dimers (ferromagnetic or antiferromagnetic) is also investigated. Chemically inhomogeneous Fe-Co atomic dimers are considered as the smallest (ordered) building block of a Fe$_{50}$Co$_{50}$ alloy \cite{lu2024}. The damping anisotropies by a rotation of the atomic spins are studied for magnetic adatoms and dimers, keeping the collinear magnetic order in the latter case.

The paper is organized as follows. In Sec.~\ref{sec_damping} we briefly summarize the computational method of the Gilbert damping within the linear response theory of exchange torque correlation as implemented in the real-space KKR formalism. Sec.~\ref{sec_comp} details the computational parameters of the studied magnetic systems in 2D and 0D. Sec.~\ref{sec:results} reports our results on magnetic Fe and Co monolayers (Sec.~\ref{sec:res_lay}) and adatoms and dimers (Sec.~\ref{sec:res_clu}) on nonmagnetic Cu, Ag and Au substrates. We draw our conclusions in Sec.~\ref{sec:conc}.

\section{Method}
\label{sec:method}

\subsection{Linear response formalism of the Gilbert damping within the real-space KKR framework}
\label{sec_damping}

The diagonal Cartesian matrix elements of the atomic-site-nonlocal intrinsic Gilbert damping tensor between magnetic sites $j$ and $k$ can be expressed as \cite{bulkcikk}
\begin{align}
\alpha_{jk}^{\mu\mu}(\eta)=-\frac{1}{4}\left[\tilde{\alpha}_{jk}^{\mu\mu}(E_F^+,E_F^+)+\tilde{\alpha}_{jk}^{\mu\mu}(E_F^-,E_F^-)\right.\nonumber\\\left.-\tilde{\alpha}_{jk}^{\mu\mu}(E_F^+,E_F^-)-\tilde{\alpha}_{jk}^{\mu\mu}(E_F^-,E_F^+)\right],
\label{eq_alpha_eta}
\end{align}
where $\mu\in\{x,y,z\}$, $E_F^{\pm}=E_F\pm i\eta$ with the Fermi energy $E_F$ and the imaginary part $\eta$, which describes the electron band broadening \cite{lu2024}, and plays the role of the scattering rate in other damping theories \cite{Gilmore2007,Barati2014,Edwards2016,Thonig2018}. The four individual terms in Eq.~(\ref{eq_alpha_eta}) are obtained by calculating the exchange torque correlations as originally proposed by Ebert et al.~\cite{Ebert2011},
\begin{align}
\label{eq_alpha}
&\tilde{\alpha}_{jk}^{\mu\mu}(E_1,E_2)=\\ 
& \frac{2}{\pi m^{j}_\mathrm{s}}  
\mathrm{Tr}\left(\underline{T}^{j}_{\mu}(E_1,E_2)\underline{\tau}^{jk}(E_2)\underline{T}^{k}_{\mu}(E_2,E_1)\underline{\tau}^{kj}(E_1)\right)\nonumber
\end{align}
with all combinations of $E_{1,2}\in\{E_F^+,E_F^-\}$. Here, $m^{j}_\mathrm{s}$ is the spin moment of site $j$, $\underline{T}^{j}_{\mu}$ is the matrix of the torque operator at site $j$, and $\underline{\tau}^{jk}$ is the matrix of the scattering path operator connecting sites $j$ and $k$.
The latter two quantities are given in angular momentum representation ($\Lambda,\Lambda'\in\{1,...,2(\ell_\mathrm{max}+1)^2\}$, where $\ell_\mathrm{max}$ is the angular momentum cutoff for describing the electron scattering): $\underline{T}^{j}_{\mu}=T^{j}_{\mu;\Lambda\Lambda'}$ and $\underline{\tau}^{jk}=\tau^{jk}_{\Lambda\Lambda'}$, respectively. 
Consequently, the trace in Eq.~(\ref{eq_alpha}) describing the torque-torque correlation is evaluated in angular-momentum space.
The matrix elements of the torque operator are calculated by the following integrals within the volume of atomic cell $j$, $\Omega_j$,
\begin{align}
& T^{j}_{\mu;\Lambda\Lambda'}(E_1,E_2)=\nonumber\\
& \int_{\Omega_j} d^3r Z^{j\times}_{\Lambda}(\vec{r},E_1)\beta\Sigma_{\mu}B_{\rm xc}^j(\vec{r})Z^{j}_{\Lambda'}(\vec{r},E_2),
\label{eq_torque}
\end{align}
where $\beta$ and $\Sigma_{\mu}$ are four-by-four matrices,
\begin{align}
\beta=\left(   \begin{array}{cc} I_2 & 0 \\ 0 & -I_2  \end{array} \right)
\qquad
\Sigma_\mu=\left(   \begin{array}{cc} \sigma_\mu & 0 \\ 0 & \sigma_\mu  \end{array} \right) \; ,
\end{align}
with the two-by-two unit matrix $I_2$ and the Pauli matrices $\sigma_{\mu}$, and $B_{\rm xc}^j(\vec{r})$ is the exchange-correlation field at site $j$ in the local spin density approximation (LSDA). $Z^{j}_{\Lambda}(\vec{r})$ are right-hand side regular solutions of the single-site Dirac equation and the superscript $\times$ denotes complex conjugation restricted to the spinor spherical harmonics only \cite{Ebert2011}.

Note that Eq.~(\ref{eq_alpha}) corresponds to the Kubo-Greenwood formula for the exchange torque correlation as a Fermi surface quantity valid for the diagonal ($\mu\mu$) elements of the Gilbert tensor only. The treatment of the off-diagonal ($\mu\ne\nu$) tensor elements would require the usage of the Kubo-Bastin formula \cite{Bastin1971,Bonbien2020}, where Fermi surface and Fermi sea terms also contribute, which is not implemented in our RS-KKR code.

The multiple-scattering of electrons in a finite cluster of $N_C$ atoms embedded into a layered 2D translation-invariant host is accounted for by \cite{Lazarovits2002}
\begin{align}
\boldsymbol{\tau}_{\mathrm{C}}(E)=\boldsymbol{\tau}_{\mathrm{H}}(E)\left[\mathbf{I}-(\mathbf{t}_{\mathrm{H}}^{-1}(E)-\mathbf{t}_{\mathrm{C}}^{-1}(E))\boldsymbol{\tau}_{\mathrm{H}}(E)\right]^{-1},
\label{eq_embedding}
\end{align}
where $\boldsymbol{\tau}_{\mathrm{C}}(E)$ and $\boldsymbol{\tau}_{\mathrm{H}}(E)$ are the scattering path operator matrices of the embedded atomic cluster and the host, respectively; $\mathbf{t}_{\mathrm{C}}(E)$ and $\mathbf{t}_{\mathrm{H}}(E)$ are the corresponding single-site scattering matrices, all in a combined atomic site ($j,k\in\{1,...,N_C\}$) and angular momentum ($\Lambda,\Lambda'\in\{1,...,2(\ell_\mathrm{max}+1)^2\}$) representation: $\boldsymbol{\tau}_{\mathrm{C}/\mathrm{H}}(E)=\{\underline{\tau}^{jk}_{\mathrm{C}/\mathrm{H}}(E) \}=\{\tau^{jk}_{\mathrm{C}/\mathrm{H};\Lambda\Lambda'}(E)\}$ and $\mathbf{t}_{\mathrm{C}/\mathrm{H}}(E)=\{t^{j}_{\mathrm{C}/\mathrm{H};\Lambda\Lambda'}(E)\delta_{jk}\}$, at the relevant energies $E=E_F^\pm$.

Using the above formalism, $\alpha_{jk}^{xx}(\eta)$, $\alpha_{jk}^{yy}(\eta)$ and $\alpha_{jk}^{zz}(\eta)$ are calculated exclusively taking magnetic sites for $j$ and $k$ in collinear magnetic configurations, where $j=k$ denote on-site and $j\ne k$ denote inter-site damping quantities. Note that the atomic-site-nonlocal Gilbert damping is generally not symmetric in the atomic site indices, $\alpha_{jk}^{\mu\mu}\neq\alpha_{kj}^{\mu\mu}$, but
\begin{align}
\label{eq:alpha_kj-jk}
m^{k}_\mathrm{s}\alpha_{kj}^{\mu\mu}=m^j_\mathrm{s}\alpha_{jk}^{\mu\mu}
\end{align}
is always true as implied by Eq.~(\ref{eq_alpha}). This is relevant in the present work for the supported mixed Fe-Co atomic dimers: $m^\mathrm{Fe}_\mathrm{s}\alpha_{\mathrm{Fe}-\mathrm{Co}}^{\mu\mu}=m^\mathrm{Co}_\mathrm{s}\alpha_{\mathrm{Co}-\mathrm{Fe}}^{\mu\mu}$. On the other hand, in an ideal ferromagnetic atomic layer with a single chemical component, $\alpha_{jk}^{\mu\mu}=\alpha_{kj}^{\mu\mu}$ applies because $m^{j}_\mathrm{s}=m^{k}_\mathrm{s}=m_\mathrm{s}$ for any pair of atomic sites.

\subsection{Computational details of the studied physical systems}
\label{sec_comp}

First, we consider Fe and Co monolayers on face-centered-cubic (fcc) Cu, Ag and Au substrates in (001) and (111) crystallographic orientations.
The bulk lattice constants of the substrates are chosen as $a_\mathrm{3D}^\mathrm{Cu}=3.615$\,\AA, $a_\mathrm{3D}^\mathrm{Ag}=4.086$\,\AA, and $a_\mathrm{3D}^\mathrm{Au}=4.079$\,\AA. Based on these values, the atomic interlayer distances in the bulk are: $a_\perp^{\mathrm{Cu(001)}}=1.807$\,\AA, $a_\perp^{\mathrm{Cu(111)}}=2.087$\,\AA, $a_\perp^{\mathrm{Ag(001)}}=2.043$\,\AA, $a_\perp^{\mathrm{Ag(111)}}=2.359$\,\AA, $a_\perp^{\mathrm{Au(001)}}=2.039$\,\AA, and $a_\perp^{\mathrm{Au(111)}}=2.355$\,\AA. The in-plane lattice constants are: $a_0^\mathrm{Cu}=2.556$\,\AA, $a_0^\mathrm{Ag}=2.889$\,\AA, and $a_0^\mathrm{Au}=2.884$\,\AA.

In our previous work \cite{bulkcikk} we showed that geometric relaxations of the atomic layers at the surface of magnetic materials affect remarkably the damping parameters in these layers. Such geometric relaxations have to be taken into account in the vicinity of the magnetic monolayers on nonmagnetic substrates as well. We used the Vienna Ab-initio Simulation Package (VASP) \cite{vasp} within LSDA \cite{Ceperley1980} to determine the optimized layered geometries, where the magnetic layer and the two neighboring (uppermost) layers of the substrate are allowed to relax in the surface-normal direction assuming pseudomorphic growth on the substrates.
The obtained relaxation values together with the atomic spin moments of the ferromagnetic monolayers are presented in Table \ref{tab:relaxations}.
We find significant inward relaxations of the magnetic layers in each case, which are systematically increasing with the increasing atomic number in the Cu-Ag-Au series. As expected, the spin moments in the surface magnetic monolayers are increased compared to their bulk values.

\begin{table}[h]
    \centering
     \caption{Calculated relative atomic layer relaxations (measured in \%) with respect to the bulk interlayer distance of the substrate 
     in the given crystallographic orientation, and atomic spin moments ($m_\mathrm{s}$) of Fe and Co monolayers on the considered substrates. M and X$_i$ denote the magnetic layer and the $i$th substrate layer measured from the substrate interface layer, respectively.}
    \label{tab:relaxations}
    \begin{tabular}{crrrc}\hline\hline
         & M--X$_1$ & X$_1$--X$_2$ & X$_2$--X$_3$ &$m_\mathrm{s}$\,($\mu_B$)\\\hline
     Fe/Cu(001) & -6.9 & -6.0 & -6.4 & 2.74\\
     Fe/Ag(001) & -18.1 & -3.5 & -4.1 & 3.19\\
     Fe/Au(001) & -20.6 & 0.9  & -1.0 & 3.27\\
     Co/Cu(001) & -9.5 & -5.9 & -6.8 & 1.80\\
     Co/Ag(001) & -22.2 & -3.4 & -4.2 & 2.12\\
     Co/Au(001) & -24.3 & 0.5 & -0.9 & 2.19\\
     Fe/Cu(111) & -3.9 & -4.1 & -3.8 & 2.55\\
     Fe/Ag(111) & -12.2 & -2.5 & -2.5 & 3.10\\
     Fe/Au(111) & -14.4 & 0.4 & -0.1 & 3.11\\     
     Co/Cu(111) & -6.3 & -4.3 & -4.0 & 1.69\\
     Co/Ag(111) & -14.9 & -2.7 & -2.4 & 2.04\\
     Co/Au(111) & -16.4 & -0.3 & -0.4 & 2.06\\ \hline\hline
    \end{tabular}
\end{table}

For defining the coordinate systems, in case of (001)-oriented surfaces the crystallographic directions $[\overline{1}10]$, $[\overline{1}\overline{1}0]$ and $[001]$ are chosen as axis $x$, $y$ and $z$, respectively, and for the (111)-oriented surfaces, the crystallographic directions $[\overline{1}10]$, $[11\overline{2}]$ and $[111]$ correspond to axis $x$, $y$ and $z$, respectively.

In case of the magnetic monolayers the diagonal Cartesian elements ($xx$, $yy$ and $zz$) of the site-dependent Gilbert damping tensor are cast as the on-site (denoted by "00"), first neighbor ("01"), second neighbor ("02") etc. quantities, all sites taken from the magnetic monolayer. Note that atomic site "0" can be arbitrarily chosen in the magnetic monolayer due to its translational symmetry. For the sake of comparability, in the presentation of the results the averages of the $xx$ components of all first ("01") or second ("02") neighbors are shown. Moreover, the effective (total) damping is calculated and analyzed, which is analogously defined as in the bulk case \cite{bulkcikk},
\begin{align}
\alpha_{\mathrm{tot}}^{\mu\mu}=
\sum\limits_{j=0}^\infty\alpha_{0j}^{\mu\mu}\approx\sum\limits_{r_{0j}\leq r_{\mathrm{max}}}\alpha_{0j}^{\mu\mu}
\label{eq_alpha_tot}
\end{align}
with an $r_\mathrm{max}$ cutoff distance measured from site "0". Evidently, $\alpha_{\mathrm{tot}}^{\mu\mu}$ equals the Fourier transform of $\alpha_{jk}^{\mu\mu}$ at $\vec{q}=0$. First, the broadening($\eta$)-dependence of the damping parameters is monitored in the range of $\eta=$1 meV to 1 eV, and then by fixing $\eta$  at 5 mRyd (0.068 eV) they are investigated as a function of the magnetization direction and substrate surface orientation. Note that for calculating the Gilbert damping in the deposited magnetic monolayers, only unperturbed host atoms form the atomic cluster, and the so-called self-embedding procedure \cite{Palotas2003} is employed, where Eq.~\eqref{eq_embedding} reduces to $\boldsymbol{\tau}_{\mathrm{C}}(E)=\boldsymbol{\tau}_{\mathrm{H}}(E)$.

Following the testing in Ref.~\cite{bulkcikk}, for obtaining $\boldsymbol{\tau}_{\mathrm{H}}(E^\pm_{\rm F})$ 45150 and 30100 $\vec{k}$ points in the irreducible parts of the Brillouin zone are used for the (001)- and (111)-oriented surfaces, respectively. The angular momentum cutoff $\ell_\mathrm{max}$ is set to 2, and for calculating $\alpha_{\mathrm{tot}}^{\mu\mu}$ in Eq.~\eqref{eq_alpha_tot} the spatial cutoff of $r_{\mathrm{max}}=20\,a_0$ is chosen in the magnetic monolayer.
Note that a circle of radius $r_{\mathrm{max}}=20\,a_0$ contains 1257 and 1459 atomic sites in the magnetic monolayer on the (001)- and (111)-oriented substrates, respectively.

Similarly to bulk magnets the non-local Gilbert damping quickly decays to zero with the distance, and can be approximated with the following function \cite{bulkcikk},
\begin{align}
 \alpha(r) \approx A \frac{\sin\left(kr+\phi_0\right)}{r^2}\exp(-\beta r) \, ,
 \label{eq:rfuggofit}
\end{align}
where $A$, $k$, $\phi_0$ and $\beta$ are the corresponding constants.
Our numerical results show that the decay rate $\beta$ strongly depends on the broadening $\eta$, whereas the wavelength~$k$ of the oscillation stays unaltered. 
Compared to the case of bulk ferromagnets in Ref.~\cite{bulkcikk}, for the magnetic monolayers we find that using the same broadening value $\eta$ the oscillation decays faster, meaning that above a radius of $r=10\,a_0$ the effect of non-local contributions becomes negligible, and $\alpha_\mathrm{tot}^{\mu\mu}$ in Eq.~\eqref{eq_alpha_tot} becomes well-converged.

Our study also includes the investigation of Gilbert damping of magnetic adatoms and small magnetic clusters in the form of atomic dimers. 
The embedded cluster KKR method \cite{Lazarovits2002} described by Eq.~\eqref{eq_embedding} is used to calculate the electronic structure of these 0D magnets on the (001)-oriented surfaces of Cu, Ag and Au. 
The layered host systems for the atomic clusters are modeled as 3 atomic layers of vacuum on the top of 9 atomic layers of the substrate material (Cu, Ag, Au), sandwiched between a semi-infinite bulk vacuum and semi-infinite bulk substrate (Cu, Ag, Au). The host geometries include the same interlayer relaxations as reported for the magnetic monolayers in Table \ref{tab:relaxations}.
The Fe and Co magnetic atoms are embedded into (fcc) atomic positions of the vacuum layer adjacent to the substrate metal. The magnetic atomic dimers are deposited along the $x$ direction on the (001)-oriented substrate surfaces. 
Self-consistent calculations of the atomic clusters are performed by including two surrounding atomic shells of perturbed host (metal substrate and vacuum) atomic sites around the magnetic atoms. Altogether, 1 magnetic site + 70 perturbed host sites are considered for the magnetic adatom systems, and 2 magnetic sites + 116 perturbed host sites for the magnetic dimers.

For the magnetic adatoms the on-site Gilbert damping ($\alpha_{11}$), and for the magnetic dimers the on-site ($\alpha_{11}$, $\alpha_{22}$) and inter-site ($\alpha_{12}$, $\alpha_{21}$) Gilbert damping values are reported according to the magnetic atomic site indices: "1" for the adatom, and "1" or "2" for the dimer. Note that here the arbitrary magnetic site notation "0" cannot be used as for the translationally invariant magnetic monolayers.
In all calculations of the Gilbert damping parameters of the 0D magnets the broadening $\eta$ is set to 0.068 eV. Though the induced magnetic moments on nonmagnetic sites are not considered in the LLG equations, Eq.~(\ref{eq:Gilbert_nonloc}), we tested the damping parameters stemming from the perturbed (originally nonmagnetic) host atoms in the self-consistently calculated atomic clusters, and found that the on-site and inter-site damping contributions involving these sites having induced magnetization are orders of magnitude smaller than on the Fe and Co sites, and thus negligible. In the following, Gilbert damping values involving magnetic Fe and/or Co atomic sites are exclusively reported.

\section{Results and discussion}\label{sec:results}

The method outlined in Sec.~\ref{sec_damping} is employed to study the Gilbert damping parameters of ferromagnetic Fe and Co  monolayers on (001)- and (111)-oriented Cu, Ag and Au substrates, as well as for Fe and Co adatoms and dimers deposited on (001)-oriented surfaces of the same nonmagnetic metals.

\subsection{Magnetic monolayers}
\label{sec:res_lay}

\begin{figure}[ht!]
  \includegraphics[width=0.48\columnwidth]{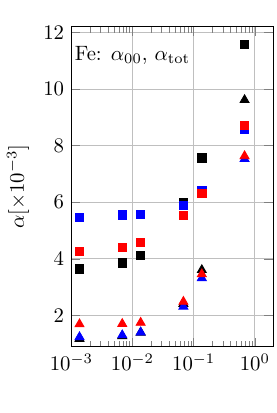}
  \includegraphics[width=0.49\columnwidth]{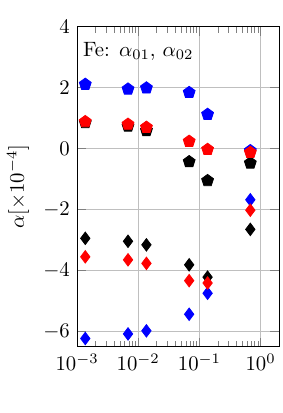}
  \\\vspace{-0.8cm}
  \includegraphics[width=0.48\columnwidth]{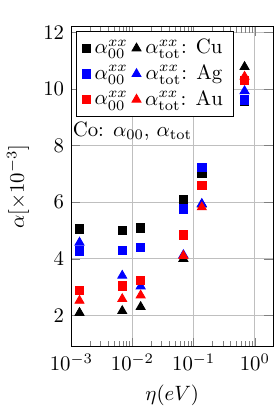}
  \includegraphics[width=0.49\columnwidth]{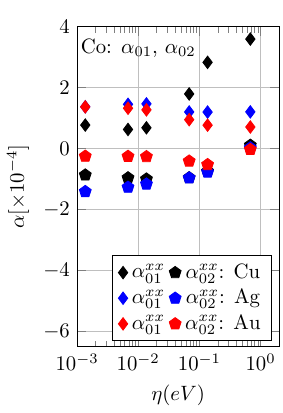}
    \caption{\label{fig:lay_broadening_dependence} Broadening dependence of Gilbert damping matrix elements in Fe and Co monolayers on Cu(001), Ag(001) and Au(001) substrates magnetized in the $\hat{\bf{z}}=[001]$ direction. Left panels: On-site ($\alpha_{00}^{xx}$) and total ($\alpha_{tot}^{xx}$) Gilbert damping tensor components as a function of the broadening $\eta$ in Fe (above) and Co monolayers (below). Right panels: Nonlocal first ($\alpha_{01}^{xx}$) and second ($\alpha_{02}^{xx}$) nearest neighbor Gilbert damping tensor components for the same systems.
    }
 \end{figure}

Figure \ref{fig:lay_broadening_dependence} shows the dependence of the $xx$ Gilbert damping matrix elements on the broadening $\eta$ for out-of-plane magnetized Fe and Co monolayers on (001)-oriented surfaces of Cu, Ag and Au.
The left panels show on-site ($\alpha_{00}$) and total ($\alpha_\mathrm{tot}$) dampings, and the right panels the first- ($\alpha_{01}$) and second-neighbor ($\alpha_{02}$) dampings.
Note that the first and second neighbor values correspond to the averages over all possible $j$th ($j=$1 or 2) neighbor sites.
The values and their qualitative $\eta$ dependence is in a very good agreement with those found in bulk ferromagnets in the given range of $\eta$ \cite{bulkcikk}.

\begin{table}[h]
    \centering
     \caption{Comparison of on-site Gilbert damping values $\alpha_{00}^{xx}=\alpha_{00}^{yy}$ in bulk Fe and Co ferromagnets, in the surface layers of bcc Fe(001) and fcc Co(001), and in Fe and Co monolayers on (001)-oriented nonmagnetic substrates with normal-to-plane magnetization direction ($\hat{\bf{z}}=[001]$), and broadening $\eta=0.068$\,eV. All damping values are muliplied by $10^{3}$.}
    \label{tab:damping_comparison}
     \begin{tabular}{lccccc}\hline\hline
     & bulk \cite{bulkcikk} & surface & \multicolumn{3}{c}{monolayers (this work)} \\ 
     &         &            & \hspace{5pt} Cu \hspace{5pt} & \hspace{5pt} Ag \hspace{5pt} & Au\\ \hline
     Fe \hspace{5pt} & 6.00 & 13.28 & 5.97 & 5.88 & 5.55 \\
     Co & 4.62 &  4.79 & 6.08 & 5.77 & 4.85 \\\hline\hline
    \end{tabular}
\end{table}

In Table \ref{tab:damping_comparison} the on-site damping values are directly compared with those calculated in bulk Fe and Co ferromagnets and in the surface atomic layer at their (001)-oriented surfaces \cite{bulkcikk}.
Interestingly, we find that the 2D magnetic layers give comparable but slightly smaller on-site damping values for Fe than in the bulk Fe, and the on-site damping is more than doubled at the surface of Fe(001). For Co we find an opposite trend, namely increased on-site damping values in the 2D magnetic layers compared to bulk Co, and the on-site damping at the surface of Co(001) is not significantly different.

Regarding the variation of the different damping tensor components with the different substrates, a clear-cut trend can hardly be observed. Since $m_\mathrm{s}$ enters the denominator of the damping formula in Eq.~\eqref{eq_alpha}, the increasing spin magnetic moment of both Fe and Co in the series of Cu-Ag-Au for the substrate (see Table \ref{tab:relaxations}) would, in principle, imply a corresponding decrease of the Gilbert damping.
Note, however, that the intrinsic damping value is a Fermi surface property in our theory, which results from a complex interplay of atomic relaxations in the vicinity of the magnetic layer and SOC effects of the substrate, all modifying the electronic structure at the magnetic sites. We will try to disentangle these effects in some extent. The effect of the atomic relaxations on the damping is systematically studied for the magnetic adatoms in Sec.~\ref{sec:res_clu}. The effect of the SOC on the damping in the magnetic monolayers is discussed in this section later on.

Let us focus now on the effect of the magnetization direction on the damping components in the magnetic monolayers on the (001)-oriented surfaces. Due to the broken symmetry we expect a significant change in the damping in the in-plane magnetized case as compared to the out-of-plane magnetization. For in-plane $\hat{\bf{x}}=[\overline{1}10]$ magnetization all three diagonal elements of the damping tensor are different, while for out-of-plane $\hat{\bf{z}}=[001]$ magnetization the two transversal diagonal components of the on-site and total Gilbert dampings are equal because of the $C_{4v}$ symmetry of the system: $\alpha_{00/\mathrm{tot}}^{xx}(\hat{\bf{z}})=\alpha_{00/\mathrm{tot}}^{yy}(\hat{\bf{z}})$.

\begin{table}[h!]
    \centering
     \caption{On-site and total damping tensor components (multiplied by $10^3$) for different magnetization directions in Fe and Co monolayers on (001)-oriented substrates. $\hat{\bf{z}}$ and $\hat{\bf{x}}$ respectively correspond to $[001]$ and $[\overline{1}10]$ magnetization directions. The broadening is $\eta=0.068$\,eV.}
    \label{tab:lay_bz-bx}
    \begin{tabular}{ccccccc}\hline\hline
        $    $ & $\alpha_{00}^{yy}(\hat{\bf{z}})$ & $\alpha_{00}^{yy}(\hat{\bf{x}})$ & $\alpha_{00}^{zz}(\hat{\bf{x}})$ & $\alpha_\mathrm{tot}^{yy}(\hat{\bf{z}})$ & $\alpha_\mathrm{tot}^{yy}(\hat{\bf{x}})$ & $\alpha_\mathrm{tot}^{zz}(\hat{\bf{x}})$ \\\hline
     Fe/Cu & 5.97 & 5.98 & 5.96 & 2.41 & 2.41 & 2.40 \\
     Fe/Ag & 5.88 & 5.83 & 5.77 & 2.31 & 2.30 & 2.27 \\
     Fe/Au & 5.55 & 5.62 & 5.62 & 2.49 & 2.58 & 2.36 \\
     Co/Cu & 6.08 & 5.98 & 5.90 & 4.01 & 3.94 & 3.75 \\
     Co/Ag & 5.77 & 5.73 & 5.61 & 4.14 & 4.08 & 3.82 \\ 
     Co/Au & 4.85 & 4.85 & 4.79 & 4.11 & 4.03 & 3.77 \\
     \hline\hline
    \end{tabular}
\end{table}

Table \ref{tab:lay_bz-bx} shows the comparison of $\alpha^{yy}(\hat{\bf{z}})$, $\alpha^{yy}(\hat{\bf{x}})$ and $\alpha^{zz}(\hat{\bf{x}})$ for the on-site and total dampings.
The difference between the $\hat{\bf{z}}$ and $\hat{\bf{x}}$ magnetization directions can be regarded as the anisotropy of the Gilbert damping.
The anisotropy of the on-site damping is zero in bulk magnets, and it is about $0.1-0.5\%$ in their (001) surface layers \cite{bulkcikk}.
Table \ref{tab:lay_bz-bx} shows generally larger damping anisotropies for the magnetic monolayers on nonmagnetic substrates, in the range of $0-3.1\%$ for $\alpha_{00}$ (maximal for Co/Cu), and in the range of $0-9\%$ for $\alpha_\mathrm{tot}$ (maximal for Co/Au). The anisotropies of the total damping are larger for the Co monolayer than for the Fe monolayer, but such a clear trend is not observed for the anisotropy of the on-site damping.

\begin{figure}[ht!]
  \includegraphics[width=0.405\columnwidth]{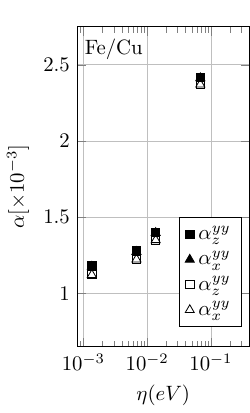} \hspace{-0.4cm}
  \includegraphics[width=0.392\columnwidth]{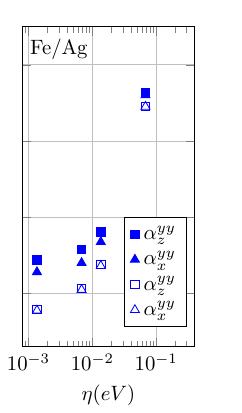} \hspace{-1.35cm}
  \includegraphics[width=0.356\columnwidth]{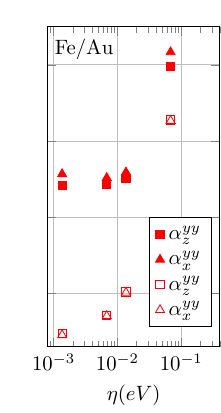}
    \caption{\label{fig:soc} Broadening dependence of the $yy$ component of the effective (total) damping $\alpha_\mathrm{tot}^{yy}$ in Fe monolayers on Cu(001), Ag(001) and Au(001) substrates for in-plane ($\alpha_{x}^{yy}=\alpha_{\mathrm{tot}}^{yy}(\hat{\bf{x}})$) and normal-to-plane ($\alpha_{z}^{yy}=\alpha_{\mathrm{tot}}^{yy}(\hat{\bf{z}})$) magnetizations. The full and empty symbols stand for calculations when spin-orbit coupling (SOC) is included or excluded, respectively. Note that for the latter case (SOC=0) $\alpha_{\mathrm{tot}}^{yy}(\hat{\bf{x}})=\alpha_{\mathrm{tot}}^{zz}(\hat{\bf{x}})$. 
}
\end{figure}

The damping anisotropy also depends on the broadening $\eta$. This is demonstrated in Figure \ref{fig:soc}, which shows the $yy$ component of the total Gilbert damping $\alpha_\mathrm{tot}^{yy}$ for the in-plane ($\hat{\bf{x}}$, triangles) and normal-to-plane ($\hat{\bf{z}}$, squares) magnetized cases for the Fe monolayers on the different (001)-oriented substrate surfaces for a broadening range with values of $\eta\le 0.068$ eV. The damping anisotropy (difference between the full symbols at a fixed $\eta$ in Fig.~\ref{fig:soc}) is very small in the Fe monolayer on Cu for all considered $\eta$ values, and it is enhanced in case of the Ag and Au substrates, which suggests that this finding is related to the increased SOC in the latter two substrate materials. To test this we excluded the SOC in the next set of our calculations.

The empty symbols in Fig.~\ref{fig:soc} represent calculated total damping values without SOC. First, we observe that the damping values themselves are lowered in the absence of SOC regardless of the direction of the magnetic moments (compare empty and full symbols at a fixed $\eta$ in Fig.~\ref{fig:soc}). This is in line with the finding for bulk ferromagnets, where the SOC-originated contribution to the Gilbert damping becomes more important as the broadening $\eta$ goes toward zero, since without SOC the damping also approaches zero \cite{bulkcikk}. Moreover, this lowering of the damping values is significantly increased going from Cu through Ag toward the Au substrate, which perfectly aligns with the increased SOC in the substrates in the same order of the chemical elements. Our next finding based on Fig.~\ref{fig:soc} is that without SOC the damping anisotropy completely vanishes (compare empty square and triangle symbols at a fixed $\eta$). Similarly, without SOC the $\alpha_\mathrm{tot}^{zz}(\hat{\bf{x}})$-$\alpha_\mathrm{tot}^{yy}(\hat{\bf{x}})$ difference, i.e., the anisotropy of the transversal components of the Gilbert damping tensor also vanishes, and as such this effect can also be attributed to the SOC. Table \ref{tab:lay_bz-bx} shows further clear evidence that the anisotropy of the transversal components of the Gilbert damping tensor at in-plane magnetization $\hat{\bf{x}}$ increases going from Cu toward Au, i.e., with increasing SOC of the substrate material.

\begin{table}[h!]
    \centering
     \caption{Dependence of $xx$ damping components (multiplied by $10^3$) of Fe and Co monolayers on the surface orientation of the substrates: comparison of (001)- and (111)-oriented surfaces with $\hat{\bf{z}}=[001]$ and $\hat{\bf{z}}=[111]$ magnetization direction, respectively. The broadening is $\eta=0.068$\,eV. Note that $\alpha^{xx}_{01}$ corresponds to the averaged value taking all first neighbors.}
    \label{tab:lay_100-111}
    \begin{tabular}{lcrrrcrrr}\hline\hline
         & &  & (001) & & &  & (111) & \\
         & & $\alpha^{xx}_{00}$ & $\alpha^{xx}_{01}$ & $\alpha^{xx}_{\mathrm{tot}}$ & & $\alpha^{xx}_{00}$ & $\alpha^{xx}_{01}$ & $\alpha^{xx}_{\mathrm{tot}}$ \\ \hline
     Fe/Cu & & 5.97 & -0.39 & 2.41 & & 11.2 & -0.91 & 2.65 \\
     Fe/Ag & & 5.88 & -0.56 & 2.31 & & 7.34 & -0.69 & 2.20 \\
     Fe/Au & & 5.55 & -0.43 & 2.49 & & 7.11 & -0.65 & 2.58 \\
     Co/Cu & & 6.08 & 0.12 & 4.01 & & 6.95 & 0.04 & 3.91 \\
     Co/Ag & & 5.77 & 0.18 & 4.14 & & 5.99 & -0.12 & 3.67 \\
     Co/Au & & 4.85 & 0.09 & 4.11 & & 5.97 & -0.02 & 4.26 \\ \hline\hline
    \end{tabular}
\end{table}

Table \ref{tab:lay_100-111} shows the comparison of Gilbert damping parameters of Fe and Co magnetic monolayers on (001)- and (111)-oriented nonmagnetic substrates.
For all cases the on-site damping component $\alpha^{xx}_{00}$ is larger for the (111)-oriented surfaces than for (001) and this enhancement is more pronounced for the Fe layers. The difference in the total damping $\alpha^{xx}_\mathrm{tot}$ between (001)- and (111)-oriented substrates is much smaller than for the on-site damping, which implies that the inter-site damping contributions almost compensate the rather large difference in the on-site damping. The presented first neighbor damping data agree with this assumption for the Fe monolayers, where larger negative $\alpha^{xx}_{01}$ values are obtained for the (111) case. For Co monolayers $\alpha^{xx}_{01}$ is much smaller in absolute value than for Fe, suggesting that farther neighbor non-local dampings have more dominant contributions in case of the Co monolayer. Note that the first neighbors are exactly at the same distance in the plane of the magnetic monolayer for both substrate surface orientations, thus the difference in the first neighbor dampings between the different surface orientations can be attributed to different local environment and symmetry.

\subsection{Magnetic adatoms}
\label{sec:res_clu}

Turning to the single magnetic adatoms, first we performed a systematic investigation of the (only existing) on-site damping parameter depending on the vertical position of Fe and Co adatoms deposited on Cu(001), Ag(001) and Au(001) surfaces.
The vertical distance between the adatom and the uppermost substrate atomic layer is varied in the range of $d_{\perp}=0.75-0.95\,a_\perp$ with $0.05\,a_\perp$ steps, where $a_\perp$ denotes the interlayer distance measured in the bulk substrate. As primary results of the self-consistent calculations we note that the spin moments of the adatom decrease monotonically with the decrease of the vertical adatom-substrate distance. This can be explained by the increased hybridization of the magnetic atom and the substrate at smaller distances. Due to the same effect, the density of states of the adatom at the Fermi level denoted by DOS($E_\mathrm{F}$) increases nearly linearly with increasing vertical adatom-substrate distance. This makes possible to interpret the variation of the damping parameter against the vertical position of the adatom as the variation against the DOS($E_\mathrm{F}$) of the adatom.

\begin{figure}[ht!]\centering
  \includegraphics[width=0.99\columnwidth]{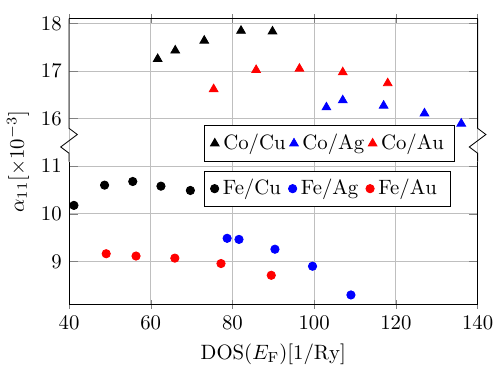}
  \includegraphics[width=0.99\columnwidth]{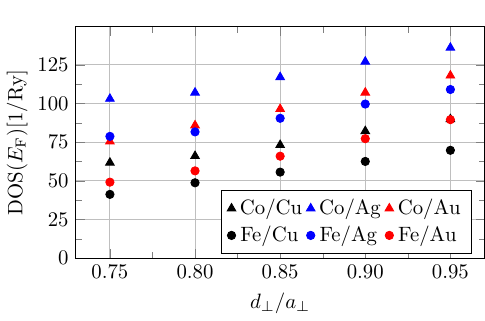}
  \includegraphics[width=0.99\columnwidth]{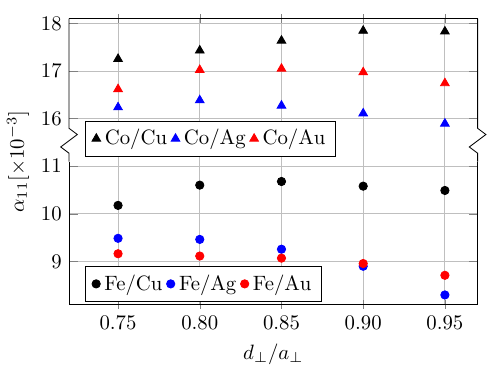}
    \caption{\label{fig:adatom_relax} Evolution of the on-site Gilbert damping $\alpha_{11}=\alpha_{11}^{xx}=\alpha_{11}^{yy}$ as a function of the electron density of states of the adatom at the Fermi level, DOS($\mathrm{E_F}$) (top) and as a function of the vertical adatom-substrate relative distance $d_{\perp}/a_{\perp}$ (bottom), for Co and Fe adatoms on the (001)-oriented surfaces of Cu, Ag and Au with $\hat{\bf{z}}=[001]$ magnetization direction. The broadening is $\eta=0.068$\,eV. Top and bottom: In each set of five points of the same symbol and same color the vertical adatom-substrate distance increases from the left (0.75$a_\perp$) to the right (0.95$a_\perp$) in steps of 0.05$a_\perp$ ($a_\perp$ denotes the interlayer distance in the bulk substrate) with a concomitant nearly linear increase of DOS($\mathrm{E_F}$), as shown in the middle panel.}
 \end{figure}

Figure \ref{fig:adatom_relax} presents the calculated on-site damping values of the magnetic adatoms as a function of DOS($E_\mathrm{F}$) (top panel) and as a function of the vertical adatom-substrate distance (bottom panel). The DOS($E_\mathrm{F}$) as  a function of the vertical adatom-substrate distance is presented in the middle panel. From these subfigures it can be inferred that the damping does not show a similar monotonic dependence on the vertical adatom-substrate distance as the adatoms' spin moments or DOS($E_\mathrm{F}$) do, instead, each magnetic atom(Fe,Co)-substrate(Cu,Ag,Au) pair has an optimal vertical distance, where the damping value is maximal. Regarding this vertical adatom position, the smallest deviation from the bulk interlayer distance ($a_\perp$) is found for the Cu substrate. Interestingly, these vertical distances providing maximal dampings tend to be around those values that were obtained for the magnetic monolayers by performing atomic geometry optimizations using VASP (see M-X$_1$ in Table \ref{tab:relaxations}). Note though that in the VASP-based relaxation not only the adatom-surface distance but also some substrate interlayer distances were optimized (see X$_i$-X$_j$ in Table \ref{tab:relaxations}).

Regarding the values of the Gilbert damping in Fig.~\ref{fig:adatom_relax} we find that Co adatoms always have larger damping values (by a factor of  about 2) than Fe adatoms at the same vertical position, in excellent agreement with Ref.~\cite{Brinker2022} considering a Au(111) substrate. This can be contrasted with the results found in bulk Fe and Co and their (001)-oriented surfaces \cite{bulkcikk} (see also Table \ref{tab:damping_comparison}), as well as with the results of the (111)-oriented monolayers (Table \ref{tab:lay_100-111}), where an opposite relation was found for the Fe and Co on-site dampings. The larger damping values of the Co adatoms correlate well with their larger DOS($E_\mathrm{F}$) as compared to the Fe adatoms at the same vertical position and same substrate, compare the triangle (Co) and circle (Fe) symbols of the same color in Fig.~\ref{fig:adatom_relax}.
In the rest of the paper we use the VASP-optimized (magnetic monolayer) geometries presented in Table \ref{tab:relaxations} for the magnetic adatoms and dimers.

\begin{figure}[ht!]
  \includegraphics[width=0.99\columnwidth]{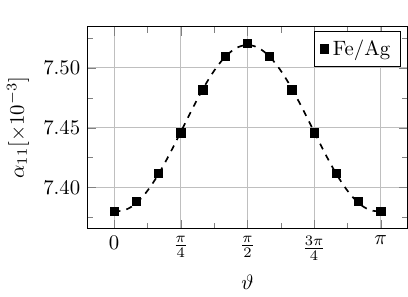}
    \caption{\label{fig:adatom_anisotropy} Calculated on-site damping of an Fe adatom on a Ag(001) surface as a function of the vertical angle $\vartheta$ of the magnetization direction with respect to the $z$ (normal-to-surface) axis. Note that the magnetization direction is always in the $xz$-plane perpendicular to the $y=[\overline{1}\overline{1}0]$ axis and $\alpha_{11}$ denotes the $yy$ component of the on-site damping tensor. The broadening is $\eta=0.068$\,eV. The dashed line shows the fitted dependence on $\vartheta$ according to Eq. \eqref{eq:damping_anis}.}
 \end{figure}

As the next step we investigate the anisotropy of the Gilbert damping with respect to the rotation of the spin moment of the magnetic adatoms around the $y=[\overline{1}\overline{1}0]$ crystallographic direction. In this case the $yy$ component of the Gilbert damping tensor is always a transversal component, thus its change can be related to the direction (or rotational angle $\vartheta$) of the spin moment. As an illustration, $\alpha_{11}^{yy}(\vartheta)$ of an Fe adatom on a Ag(001) substrate is shown in Figure \ref{fig:adatom_anisotropy}. First of all, the obtained on-site damping values are found to be larger than for the Fe monolayer on Ag(001), see Table \ref{tab:lay_bz-bx}. Moreover, we find that the magnetization-direction dependence of on-site damping can be well described with the following function,
\begin{align}
\label{eq:damping_anis}
\alpha_{11}^{yy}(\vartheta)=\alpha_0+\alpha_2 \cos^2\vartheta+\alpha_4\cos^4\vartheta,
\end{align}
where $\alpha_0$ is the damping value at the in-plane magnetized case ($\vartheta=\pi/2$), and $\alpha_2$ and $\alpha_4$ are second and fourth order anisotropic Gilbert damping parameters, respectively.

For the case of Fe/Ag(001) the maximal damping value is obtained at $\vartheta=\pi/2$, i.e., at the in-plane magnetization, and the amplitude of the change is about 2\% of its maximal value.
This anisotropy is much smaller than the one found in Ref.~\cite{Brinker2022} for an Fe adatom on Au(111) substrate, where the anisotropy of the damping-like term was identified at nearly 50\% with damping values 2 orders of magnitude larger than in our case. 
These differences might be explained by a different effective broadening value $\eta$ value in the cited reference, since we showed for the magnetic monolayers that the damping anisotropy is a SOC-induced effect and, thus, it is sensitive to the particular choice of $\eta$ (see Fig.~\ref{fig:soc}).

\begin{table}[]
    \centering
    \caption{Spin magnetic moments ($m_\mathrm{s}$), on-site Gilbert damping parameters ($\alpha_0$), relative second ($\alpha_2/\alpha_0$) and fourth order ($\alpha_4/\alpha_0$) anisotropies, see Eq.~(\ref{eq:damping_anis}), for Fe and Co adatoms on (001)-oriented fcc substrates of Cu, Ag, and Au. The broadening is $\eta=0.068$\,eV.}
    \label{tab:anisotropy_adatom}
    \begin{tabular}{cccrr}\hline\hline
                 &   $m_\mathrm{s}(\mu_\mathrm{B})$ & $\alpha_0(\times 10^{-3})$ & $\alpha_2/\alpha_0$ & $\alpha_4/\alpha_0$\\\hline
     Fe/Cu & 3.15 & 10.80 & -0.49\% & 0.01\%\\
     Fe/Ag & 3.53 & 7.52 & -2.10\% & -0.24\%\\
     Fe/Au & 3.55 & 6.38 & -0.13\% & 0.00\%\\
     Co/Cu & 1.88 & 17.39  & -1.61\% & 0.06\%\\
     Co/Ag & 2.28 & 14.87 & -3.78\% & 1.26\%\\
     Co/Au & 2.30 & 15.29 & -1.05\% & -0.01\%\\ \hline\hline
    \end{tabular}
\end{table}

The magnitudes of the spin moments and the fitted damping parameters according to Eq.~\eqref{eq:damping_anis} are presented in Table \ref{tab:anisotropy_adatom} for all investigated adatom cases.
The spin moments show a similar trend as was identified for the monolayers, see Table \ref{tab:relaxations}. The $\alpha_0$ damping values of Co are always larger than of Fe, in agreement with Ref.~\cite{Brinker2022}, where an Au(111) substrate was taken.
Note that the anisotropy effect is, in general, smaller than the change of the overall damping parameter caused by the geometric relaxation, but still observable. The relative second order damping anisotropy $\alpha_2/\alpha_0$ is always negative and it is the largest on the Ag substrate. The relative fourth order damping anisotropy $\alpha_4/\alpha_0$ is almost negligible for Au and Cu substrates, and significantly larger for the Ag substrates.
A similar anisotropy of the magnitudes of the spin moments ($m_\mathrm{s}$) can also be observed (not shown), but this effect is even smaller by two orders of magnitude.
Damping calculations on magnetic adatoms without SOC show that the SOC not only gives a (broadening dependent) contribution to the damping, but the observed damping anisotropy can also be related to the presence of the SOC.

\subsection{Magnetic dimers}

Magnetic atomic dimers provide a platform for studying the Gilbert damping properties for more complex physical scenarios than for the single adatoms: (i) collinear ferromagnetic (FM) and antiferromagnetic (AFM) ordering between the magnetic sites can be considered, and (ii) the magnetic atomic dimer can be composed of different chemical elements. In this section we investigate both scenarios.

\subsubsection{Ferromagnetic dimers}

First, let us discuss the anisotropic Gilbert damping for FM ordering. In analogy to the adatom case described in the previous section, now the two spin moments in the dimer are rotated around the $y=[\overline{1}\overline{1}0]$ direction by angle $\vartheta$ while keeping them parallel during the rotation, thus, maintaining the FM order. Again, the $yy$ transversal component of the Gilbert damping tensor is chosen for our analysis.

\begin{figure}[ht!]
  \includegraphics[width=0.99\columnwidth]{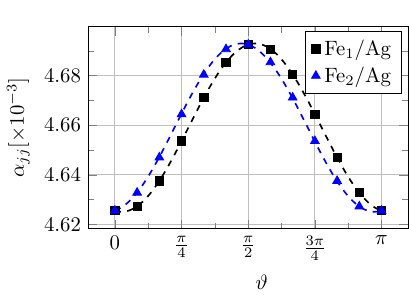}
    \caption{\label{fig:dimer_anisotropy} Calculated on-site damping parameters of an Fe-Fe FM dimer on a Ag(001) surface as a function of the vertical angle $\vartheta$ of the magnetization direction with respect to the $z$ (normal-to-surface) direction. The magnetization is always in the $xz$-plane normal to the $y=[\overline{1}\overline{1}0]$ direction. Here $\alpha_{jj}$ ($j=1,2$) denotes the $yy$ component of the on-site damping tensors: the black squares and blue triangles show the damping of the first ($\alpha_{11}^{yy}$) and second ($\alpha_{22}^{yy}$) Fe atom in the FM dimer, respectively. The dashed line shows the fitted dependence on $\vartheta$ according to Eqs.~\eqref{eq:damping_anis_dimer1} and \eqref{eq:damping_anis_dimer2}. The broadening is $\eta=0.068$\,eV.}
 \end{figure}

Figure \ref{fig:dimer_anisotropy} shows the on-site damping values $\alpha_{11}^{yy}(\vartheta)$ and $\alpha_{22}^{yy}(\vartheta)$ for the two Fe atoms in an FM Fe dimer on Ag(001). The obtained on-site damping values are found to be smaller than for the Fe adatom on Ag(001) (see Table \ref{tab:anisotropy_adatom}), and also smaller than for the Fe monolayer on Ag(001) (see Table \ref{tab:lay_bz-bx}). Quite importantly, the two on-site damping curves do not overlap and  the damping values on the two Fe atoms are equal for in-plane ($\vartheta=\pi/2$) and normal-to-plane ($\vartheta=0,\pi$) magnetic configurations only. This is due to the fact that the mirror plane intersecting the dimer's line at its center requires that $\alpha_{22}^{yy}(\vartheta)=\alpha_{11}^{yy}(\pi-\vartheta)$, which should be compared to the restriction for the adatom, $\alpha_{11}^{yy}(\vartheta)=\alpha_{11}^{yy}(\pi-\vartheta)$. The on-site damping values in the dimer, thus, follow a similar $\vartheta$-dependence as found for the adatoms, but due to the identified lower symmetry the extrema are no longer restricted to the in-plane and normal-to-plane magnetic configurations, and a new parameter, $\vartheta_0$ angle, must be included into the fitting of the calculated data points,
\begin{align}
\label{eq:damping_anis_dimer1}
\alpha_{11}^{yy}(\vartheta)=\alpha_0+\alpha_2 \cos^2\left(\vartheta+\vartheta_0\right)+\alpha_4 \cos^4\left(\vartheta+\vartheta_0\right) \, , \\
\label{eq:damping_anis_dimer2}
 \alpha_{22}^{yy}(\vartheta)=\alpha_0+\alpha_2 \cos^2\left(\vartheta-\vartheta_0\right)+\alpha_4 \cos^4\left(\vartheta-\vartheta_0\right) \, .
\end{align}
Here, $\alpha_0$ is the maximal damping value, and, as before, $\alpha_2$ and $\alpha_4$ are second and fourth order anisotropic Gilbert damping parameters, respectively.

For the case of the FM Fe dimer on Ag(001) shown in Fig.~\ref{fig:dimer_anisotropy} the maximum damping value belongs to such configurations, where the spin moments are tilted by a small angle of $\vartheta_0=\pm 0.0254 \, \pi$ ($4.57^\circ$) with respect to the in-plane magnetic direction ($\vartheta=\pi/2$).

\begin{table}[h!]
    \centering
    \caption{Spin magnetic moments ($m_\mathrm{s}$) and on-site Gilbert damping parameters ($\alpha_0$), and relative second and fourth order anisotropies ($\alpha_2/\alpha_0$ and $\alpha_4/\alpha_0$) according to Eqs.~(\ref{eq:damping_anis_dimer1}) and Eqs.~(\ref{eq:damping_anis_dimer2}) for ferromagnetic (FM) atomic dimers (deposited along the $x$ direction) of Fe and Co on (001)-oriented fcc substrates of Cu, Ag, and Au. Inter-site $\alpha_{12}^{yy}$ Gilbert damping values are also reported for $\hat{\bf{z}}=[001]$, i.e., $\vartheta=0$. The broadening is $\eta=0.068$\,eV.}
    \label{tab:anisotropy_fm_dimer}
    \begin{tabular}{cccrrc}\hline\hline
                 &$m_\mathrm{s}(\mu_\mathrm{B})$  &$\alpha_0(\times 10^{-3})$ &$\alpha_2/\alpha_0$& $\alpha_4/\alpha_0$ &$\alpha_{12}^{yy}(\times 10^{-3})$\\\hline
     Fe/Cu & 3.04 & 5.35 & 1.41\% & 0.00\% & -0.60\\
     Fe/Ag & 3.42 & 4.69 & -1.44\% & 0.00\% & -0.14\\
     Fe/Au & 3.48 & 4.81 & -0.62\% & 0.01\% & -0.50\\
     Co/Cu & 1.88 & 9.95 & -1.13\% & 0.16\% & -0.35\\
     Co/Ag & 2.25 & 8.87 & 1.57\% & -0.12\% & -0.47\\
     Co/Au & 2.30 & 10.59 & 2.86\% & 0.11\% & -0.87\\ \hline\hline
    \end{tabular}
\end{table}

The magnitudes of the spin moments, the fitted damping parameters according to Eqs.~\eqref{eq:damping_anis_dimer1} and \eqref{eq:damping_anis_dimer2} and the inter-site damping values $\alpha_{12}^{yy}$ are presented in Table \ref{tab:anisotropy_fm_dimer} for the FM Fe and Co atomic dimers.
The spin moments show a similar trend as identified for the monolayers in Table \ref{tab:relaxations} and for the adatoms in Table \ref{tab:anisotropy_adatom}, and their magnitude is slightly smaller in places than for the single adatoms. The $\alpha_0$ damping values of Co are always larger than of Fe, but in all cases these values are considerably reduced than found for the single magnetic adatoms (see Table \ref{tab:anisotropy_adatom}), most significantly in case of the Cu substrate. The inter-site dampings $\alpha_{12}^{yy}$ are by an order of magnitude smaller compared to the on-site dampings, and their consistently negative sign means that the effective total damping is also reduced compared to the on-site value. Similarly to our results, negative symmetric inter-site damping components were identified for Fe and Co FM atomic dimers on a Au(111) substrate \cite{Brinker2022}, where the negative sign of the bilinear term dominates.
The relative second order damping anisotropies $\alpha_2/\alpha_0$ in Table \ref{tab:anisotropy_fm_dimer} are in the same range as found for the single adatoms, but their sign shows more changes. The largest value of $\alpha_2/\alpha_0$ is found for Co/Au. The relative fourth order damping anisotropy $\alpha_4/\alpha_0$ values are generally negligible.
The anisotropy is also present in the inter-site dampings on a similar scale (not shown).
The damping calculations without SOC show similar results as for the adatoms: the magnitude of the dampings decreases, the anisotropy vanishes, thus $\alpha_{11}^{yy}$, $\alpha_{22}^{yy}$ and $\alpha_{12}^{yy}$ become independent of the direction of the magnetization.


Next we investigate chemically inhomogeneous FM Fe-Co atomic dimers which can be regarded as the smallest (ordered) building block of a Fe$_{50}$Co$_{50}$ alloy \cite{lu2024}.
The optimized vertical layer relaxations of the magnetic monolayer cases reported in Table \ref{tab:relaxations} are also assumed for these dimers, placing the Fe and Co atoms at the same vertical distance from the substrate's top atomic layer. Since the optimal geometries for the Fe-Fe and Co-Co dimers are different (denoted by "Fe" and "Co", respectively), the calculations for the mixed Fe-Co dimers are performed with both geometries to be able to assess how they influence the damping properties. According to Table \ref{tab:relaxations} the "Co" geometry corresponds to a larger inward relaxation, i.e., smaller vertical distance between the substrate and the Fe-Co dimer compared to the "Fe" geometry.

\begin{table}[]
    \centering
    \caption{Spin magnetic moments ($m_\mathrm{s}^{\mathrm{Fe/Co}}$), and $yy$ Cartesian components of the on-site ($\alpha_{11}^\mathrm{Fe}$, $\alpha_{22}^\mathrm{Co}$) and inter-site ($\alpha_{12}^\mathrm{Fe}$, $\alpha_{21}^\mathrm{Co}$) Gilbert damping tensor elements for normal-to-plane ($\hat{\bf{z}}=[001]$) magnetized FM Fe-Co atomic dimers (deposited along the $x$ direction) on (001)-oriented fcc substrates of Cu, Ag, and Au. The broadening is $\eta=0.068$\,eV. Two types of atomic geometries with different vertical relaxations are used according to Table \ref{tab:relaxations}: "Fe" denotes the geometries used for Fe monolayers, Fe adatoms and Fe-Fe dimers, and "Co" denotes those used for Co monolayers, Co adatoms and Co-Co dimers. "Co" position is always closer to the substrate than "Fe" position.}
    \label{tab:anisotropy_vegyes_dimer}
    \begin{tabular}{cccccrrrr}\hline\hline
      && $m_\mathrm{s}^\mathrm{Fe}$ &   $m_\mathrm{s}^\mathrm{Co}$ \;&&$\alpha_{11}^\mathrm{Fe}$  &$\alpha_{22}^\mathrm{Co}$ &$\alpha_{12}^\mathrm{Fe}$  &$\alpha_{21}^\mathrm{Co}$\\
      geom. & substr. &\multicolumn{2}{c}{($\mu_B$)}&&\multicolumn{4}{c}{($\times 10^{-3}$)}\\\hline
     "Fe"&Cu & 3.11 & 1.83 && 5.24 & 8.63 & -0.08 & -0.14\\
     "Fe"&Ag & 3.46 & 2.18 && 4.13 & 9.01 & -0.03 & -0.04\\
     "Fe"&Au & 3.51 & 2.22 && 4.13 & 10.79 & -0.46 & -0.73\\
     "Co"&Cu & 3.11 & 1.83 && 5.29 & 8.87 & -0.12 & -0.20\\
     "Co"&Ag & 3.49 & 2.18 && 4.06 & 9.20 & -0.04 & -0.06\\
     "Co"&Au & 3.54 & 2.27 && 3.75 & 10.85 & -0.39 & -0.61\\
     \hline\hline
    \end{tabular}
\end{table}

Table \ref{tab:anisotropy_vegyes_dimer} reports the spin moments and the on-site and inter-site Gilbert damping values of the mixed FM Fe-Co dimers. Compared to the chemically homogeneous FM atomic dimers (Table \ref{tab:anisotropy_fm_dimer}) the spin moments of the Fe atoms are slightly larger and those of the Co atoms are slightly smaller. Focusing on the damping values, a general result is that, independently of the geometry and the substrate, the Co atom always exhibits a larger on-site damping than the Fe atom in the mixed dimer. This is in line with the reported larger effective damping in Co-centered rather than Fe-centered small atomic clusters in FeCo alloys \cite{lu2024}, albeit at the smallest Fe-Co cluster size in our dimer case. Note that the on-site damping values for Co were also found larger for the single magnetic adatoms (see Table \ref{tab:anisotropy_adatom}) and in the FM dimers (see Table \ref{tab:anisotropy_fm_dimer}). In the mixed Fe-Co dimer the on-site damping values of Fe, $\alpha_{11}^\mathrm{Fe}$ are always smaller than the corresponding maximum values $\alpha_0$ in Table \ref{tab:anisotropy_fm_dimer}. For Co this is true on the Cu substrate only, but $\alpha_{22}^\mathrm{Co}$ is enlarged on the Ag and Au substrates compared to the corresponding $\alpha_0$ values in Table \ref{tab:anisotropy_fm_dimer}. We note here that a nonmonotonic behavior of the adatom's damping parameter as a function of the vertical adatom-substrate distance was identified in Fig.~\ref{fig:adatom_relax}, thus, a general trend of the on-site damping values for the Fe-Co mixed dimers in the "Fe" and "Co" geometries cannot be established. The geometric effect is noticeable, but it is smaller than the change caused by the chemical replacement of one atom in the FM dimer. The inter-site damping values $\alpha_{12}^\mathrm{Fe}$ and $\alpha_{21}^\mathrm{Co}$ in Table \ref{tab:anisotropy_vegyes_dimer} are all negative, and they clearly demonstrate the validity of Eq.~\eqref{eq:alpha_kj-jk}.

The anisotropic nature of the Gilbert damping in the mixed Fe-Co FM dimers is similar to the homogeneous FM dimers, which is described by Eqs.~\eqref{eq:damping_anis_dimer1} and \eqref{eq:damping_anis_dimer2}: in the mixed dimer cases $\alpha_2/\alpha_0$ is about 2-4\% for Co and 1-2\% for Fe, and $\alpha_4/\alpha_0$ is never larger than $0.1\%$.
Though in FeCo thin films a giant anisotropic damping of 400\% is reported \cite{li2019}, such high values are not found for the intrinsic damping in our Fe-Co dimers.

\subsubsection{Antiferromagnetic dimers}

So far FM atomic dimers were investigated. The final part of our paper concentrates on AFM atomic dimers composed of the same transition metal, either Fe or Co.
The same atomic geometries are used as for the FM dimers, but now the spin moment vectors of the two sites are set antiparallel to each other. Since the ground state of these dimers is ferromagnetic, considering antiferromagnetic configurations for them serves as a case study of the Gilbert damping in AFM dimers.

\begin{figure}[ht!]
  \includegraphics[width=0.99\columnwidth]{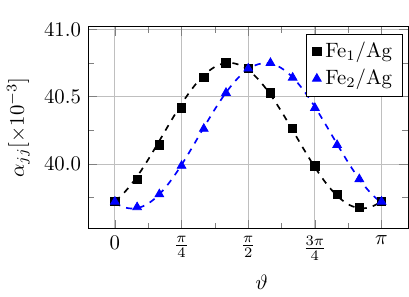}
    \caption{\label{fig:afm_anisotropy} Calculated on-site damping parameters of an AFM Fe-Fe dimer on a Ag(001) surface  as a function of the rotation angle $\vartheta$ of the magnetization around the $y=[\overline{1}\overline{1}0]$ axis, so that the spin moments always lie in the $xz$-plane. Here $\alpha_{jj}$ ($j=1,2$) denotes the $yy$ component of the on-site damping tensors: the black squares and blue triangles show the damping of the first ($\alpha_{11}^{yy}$) and second ($\alpha_{22}^{yy}$) Fe atom in the AFM dimer, respectively. The dashed line shows the fitted dependence on $\vartheta$ according to Eqs.~\eqref{eq:damping_anis_dimer1} and \eqref{eq:damping_anis_dimer2}. The broadening is $\eta=0.068$\,eV.}
\end{figure}

In particular, the damping anisotropy is studied. In analogy to the FM dimer case, the two opposite spin moments in the AFM dimer are simultaneously rotated around the $y=[\overline{1}\overline{1}0]$ direction by angle $\vartheta$, maintaining thus the AFM order. 
Figure \ref{fig:afm_anisotropy} shows the on-site damping values $\alpha_{11}^{yy}(\vartheta)$ and $\alpha_{22}^{yy}(\vartheta)$ for the two Fe atoms in an AFM Fe-Fe dimer on Ag(001). The obtained on-site damping values are found to be largely enhanced compared to the FM Fe-Fe dimer on Ag(001) (see Fig.~\ref{fig:dimer_anisotropy}), the enhancement is almost ninefold. Again, the two on-site damping curves do not overlap but they intersect each other at the in-plane ($\vartheta=\pi/2$) and normal-to-plane ($\vartheta=0,\pi$) AFM magnetic configurations. For fitting the damping curves, Eqs.~\eqref{eq:damping_anis_dimer1} and \eqref{eq:damping_anis_dimer2} are still a valid choice.

\begin{table}[h!]
    \centering
    \caption{Spin magnetic moments ($m_\mathrm{s}$) and on-site Gilbert damping parameters, $\alpha_0$, and relative second and fourth order anisotropies, $\alpha_2/\alpha_0$ and $\alpha_4/\alpha_0$, respectively, according to Eqs.~(\ref{eq:damping_anis_dimer1}) and (\ref{eq:damping_anis_dimer2}) for antiferromagnetic (AFM) atomic dimers of Fe and Co deposited along the $x$ direction on (001)-oriented fcc substrates of Cu, Ag, and Au. Inter-site $\alpha_{12}^{yy}$ Gilbert damping values are also reported for $\hat{\bf{z}}=[001]$, i.e., $\vartheta=0$. The broadening is $\eta=0.068$\,eV.}
    \label{tab:anisotropy_afm_dimer}
    \begin{tabular}{cccrrc}\hline\hline
                 &$m_\mathrm{s}(\mu_\mathrm{B})$  &$\alpha_0(\times 10^{-3})$ &$\alpha_2/\alpha_0$& $\alpha_4/\alpha_0$ &$\alpha_{12}^{yy}(\times 10^{-3})$\\\hline
     Fe/Cu & 3.01 & 55.60 & -1.38\% & -0.01\% & 23.12\\
     Fe/Ag & 3.45 & 39.68 & 2.57\% & 0.15\% & 16.76\\
     Fe/Au & 3.51 & 17.57 & 0.76\% & 0.06\% & 5.81\\     Co/Cu & 1.72 & 63.32 & -0.52\% & 0.01\% & 23.72\\
     Co/Ag & 2.23 & 43.57 & 5.00\% & 1.72\% & 15.48\\
     Co/Au & 2.28 & 33.26 & 1.86\% & 0.09\% & 10.13\\ \hline\hline
    \end{tabular}
\end{table}

The magnitudes of the spin moments, the fitted damping parameters according to Eqs.~(\ref{eq:damping_anis_dimer1}) and (\ref{eq:damping_anis_dimer2}) and the inter-site damping values $\alpha_{12}^{yy}$ are presented in Table \ref{tab:anisotropy_afm_dimer} for all investigated AFM atomic dimer cases.
The spin moments are systematically slightly smaller than for the FM dimers (see Table \ref{tab:anisotropy_fm_dimer}). The $\alpha_0$ damping values of Co are still always larger than of Fe, and the most interesting result is that in all cases the $\alpha_0$ values are considerably enhanced than found for the FM dimers (see Table \ref{tab:anisotropy_fm_dimer}), most significantly for the Cu substrate cases, where the largest $\alpha_0$ values are found, in an opposite trend for the substrate materials than for FM dimers. Overall, a 3-10 times increase of $\alpha_0$ is obtained in the AFM dimers compared to the FM cases. The obtained damping differences are in a good agreement with Ref.~\cite{Brinker2022}, where a 2-5 times increase of the (symmetric) damping was reported for Fe and Co AFM dimers in comparison to FM ones on a Au(111) surface.
A possible explanation for these findings is that the AFM ordering is energetically unfavored in comparison to the FM ordering in the considered transition metal dimers, and the spin relaxation from these AFM configurations is much faster exhibiting larger energy dissipation.

In agreement with Ref.~\cite{Brinker2022} for Fe and Co dimers on Au(111), the inter-site dampings $\alpha_{12}^{yy}$ in the AFM ordering systematically possess a positive sign as opposed to their negative sign in the FM case, see Table~\ref{tab:anisotropy_fm_dimer}, and their magnitudes are largely enhanced to 12-120 times of the corresponding FM values. Consequently, for the AFM dimers the inter-site dampings are found in the same order of magnitude as the on-site ones: from Table~\ref{tab:anisotropy_afm_dimer} one can read off non-local to on-site damping ratios, $\alpha_{12}^{yy}/\alpha_0$, in the range of 30-42\%. Such large ratios, according to our knowledge, have not yet been reported.
The relative second order damping anisotropies $\alpha_2/\alpha_0$ are in the same range as for the FM dimers. The largest value of $\alpha_2/\alpha_0$ is found for Co/Ag. The relative fourth order damping anisotropy $\alpha_4/\alpha_0$ values are generally negligible with the exception of Co/Ag.

The damping calculations without SOC show an intriguing result for the AFM dimers. Though the damping anisotropy vanishes as in all of the previous cases, the on-site and inter-site damping values themselves increase, in spite of a slight increase of the spin moments. For example, for the AFM Fe dimer on Ag,
$\alpha_{11}^{yy}(\vartheta=\pi/2)=\alpha_{22}^{yy}(\vartheta=\pi/2)=0.0407$ (see Fig.~\ref{fig:afm_anisotropy})
and $\alpha_{12}^{yy}(\vartheta=\pi/2)=\alpha_{21}^{yy}(\vartheta=\pi/2)=0.0170$ when SOC is included, while
$\alpha_{11}^{yy}(\vartheta)=\alpha_{22}^{yy}(\vartheta)=0.0478$ and $\alpha_{12}^{yy}(\vartheta)=\alpha_{21}^{yy}(\vartheta)=0.0207$ for all $\vartheta$ when SOC is switched off. These observations imply that in case of the AFM dimers the SOC has a negative damping contribution, affecting both the on-site and the inter-site dampings. The unexpected increasing trend of $\alpha_0$  obtained for the AFM dimers in the series of Au, Ag, and Cu substrates, i.e., in the order of decreasing SOC, see Table \ref{tab:anisotropy_afm_dimer}, can also be attributed to this effect.

\section{Conclusions}
\label{sec:conc}
We used a real-space \textit{ab initio} scheme based on the exchange torque correlation implemented in Ref.~\cite{bulkcikk} to calculate the Cartesian diagonal elements of the atomic-site-dependent intrinsic Gilbert damping tensor in magnets of reduced dimensions. First, we investigated ultrathin 2D monolayers of Fe and Co on Cu, Ag, and Au substrates in (001) and (111) surface orientations. The damping values are found to differ from those of the ferromagnetic 3D bulk and (001) surfaces, and the damping anisotropy is increased. Our calculations evidenced that the damping anisotropy is the result of the spin-orbit coupling, and, accordingly, a clear trend of the damping anisotropy was established with respect to the substrates. The substrate orientation significantly affected the damping values in the magnetic monolayers, however, the differences in the on-site and the inter-site contributions largely compensated each other, resulting in similar values of the total damping parameters for the (001) and (111) substrate orientations.

We also studied the Gilbert damping properties of Fe and Co adatoms and dimers on (001)-oriented substrates, and found that the Co systems systematically exhibited larger on-site damping than Fe. Moreover, the damping of single adatoms showed a non-monotonic dependence on their vertical position as quantified in terms of the adatoms' density of states at the Fermi energy. By rotating the spin moments of the adatoms and collinear ferromagnetic and antiferromagnetic dimers, we calculated second order anisotropy contributions in a range of a few percents of the damping parameters, while the fourth order anisotropy was, at least, by one order less in magnitude. For mixed Fe-Co ferromagnetic atomic dimers the on-site damping parameters are similar to their counterparts in the chemically homogeneous dimers, but, with exception of the Fe-Co dimer on Au(001) substrate, the inter-site dampings are considerably decreased in magnitude. In antiferromagnetic dimers we found a significant, three to ten times increase of the on-site Gilbert damping in comparison to the ferromagnetic dimers, while the inter-site damping is even more enhanced, resulting in non-local to on-site damping ratios of up to 42\%.

In summary, our theoretical method shows that further investigations of Gilbert damping coefficient in magnets of reduced dimensions (surfaces, interfaces, artificial nanostructures, etc.) is very promising and hopefully can bring us to a deeper understanding of spin dynamic processes at the atomic scale. We expect that our theoretical results will inspire experimentalists to develop suitable techniques for studying the magnetic damping properties in reduced dimensions.

\begin{acknowledgments}
The authors acknowledge discussions with Samir Lounis. Financial support of the National Research, Development, and Innovation (NRDI) Office of Hungary under Project No.\ FK124100, No.\ K131938, No. K142652 and ADVANCED 149745, the János Bolyai Research Scholarship of the Hungarian Academy of Sciences (Grant No. BO/292/21/11), the New National Excellence Program of the Ministry for Culture and Innovation from NRDI Fund (Grant No.\ ÚNKP-23-5-BME-12), and the Hungarian State E\"otv\"os Fellowship of the Tempus Public Foundation (Grant No.\ 2016-11) are gratefully acknowledged. Further support was provided by the Thematic Area Excellence Program of the Ministry for Culture and Innovation from NRDI Fund (Grant No.\ TKP2021-NVA-02).
\end{acknowledgments}

\end{document}